\documentclass{article}
\usepackage{fullpage,fourier,charter}
\usepackage{graphicx} 
\usepackage{amsmath}
\usepackage{mathtools}
\usepackage{amssymb}
\usepackage{authblk}
\usepackage[colorlinks=true,citecolor=blue]{hyperref}

\title{Quasi-continuum approximations for nonlinear dispersive waves in general discrete conservation laws}
\author[1]{Su Yang\thanks{Corresponding author: suyang@umass.edu}}
\affil[1]{Department of Mathematics and Statistics, University of Massachusetts Amherst, Amherst, MA 01003-4515, USA}

\date{\small\today}

\numberwithin{equation}{section}

\begin{document}

\maketitle

\begin{abstract}
    In this paper, we study a non-integrable discrete lattice model which is a variant of an integrable discretization of the standard Hopf equation. Interestingly, a direct numerical simulation of the Riemann problem associated with such a discrete lattice shows the emergence of both the dispersive shock wave (DSW) and rarefaction wave (RW). We propose two quasi-continuum models which are represented by partial differential equations (PDEs) in order to both analytically and numerically capture the features of the DSW and RW of the lattice. Accordingly, we apply the DSW fitting method to gain important insights and provide theoretical predictions on various edge features of the DSW including the edge speed and wavenumber. Meanwhile, we analytically compute the self-similar solutions of the quasi-continuum models, which serve as the approximation of the RW of the lattice. We then conduct comparisons between these numerical and analytical results to examine the performance of the approximation of the quasi-continuum models to the discrete lattice. 
\end{abstract}

\tableofcontents

\section{Introduction}

Dispersive shock wave \cite{Hoefer:2009}, which is a unstationary and dispersive wave structure, is numerically observed in a variety of mathematical physics models. Some typical examples include the famous Toda lattice \cite{BIONDINI2024134315, CHONG2024103352, toda2012theory}, the granular crystal lattice \cite{Yang_Biondini_Chong_Kevrekidis_2025, yang2025firstordercontinuummodelsnonlinear, CHONG2022133533, PhysRevE.95.062216, PhysRevLett.120.194101}, the nonlinear Schr\"odinger model \cite{mohapatra2025dambreaksdiscretenonlinear, PhysRevA.110.023304, Chandramouli_2023, PhysRevLett.100.084504}. Remarkably, besides those numerical emergences of the dispersive shock waves, they are also experimentally observed \cite{PhysRevE.75.021304, PhysRevE.80.056602, PhysRevLett.120.194101}. Importantly, the core of the DSW is the so-called periodic wave associated with the model, as it can be regarded as the slow spatial and temporal modulation on distinct parameters (e.g. wavenumber, wave amplitude) of the periodic solutions. The analysis of this particular dispersive wave structure is important. The so-called \textit{Whitham modulation theory} \cite{whitham2011linear, Abeya_2023, https://doi.org/10.1111/sapm.12651, Ablowitz_2018, PhysRevE.96.032225} provides a formal theoretical analysis, via a system of partial differential equations which are known as Whitham modulation equations, on the space-time evolution of these slowly varying and modulated parameters of the periodic waves. Moreover, the reduction of the Whitham modulation equations near both the linear and solitonic limits gives important insights on the edge features (e.g. solitonic-edge speed, linear-edge wavenumber and speed) of the dispersive shock waves. Such a reduction of the modulation system is entitled with the \textit{DSW fitting} method \cite{EL201611}. The rarefaction wave, on the other hand, is classified as a simple wave structure \cite{whitham2011linear, carretero2024nonlinear}. The dispersive model also serves as a prototypical medium for the emergence of this particular wave structure. Analytically, the rarefaction wave is described by a self-similar solution of the corresponding model. As one shall see, this work investigates both the dispersive shock and rarefaction wave of a discrete lattice model stemmed from a scalar conservation law.

The continuum scalar conservation law \cite{evans1998partial} is a dispersionless system in the sense that its linear dispersion relation has zero second derivative with respect to the wavenumber. For example, both the transport and Hopf equations in $1+1$ time-space dimension are dispersionless. Therefore, one should not expect any dispersive wave structure (e.g. DSW) to emerge in these systems. Recently, the work \cite{https://doi.org/10.1111/sapm.12767} proposed a discrete lattice model which serves as a discrete and dispersive regularization of the standard continuum Hopf equation:
\begin{equation}\label{eq: first discretization scheme}
    \frac{du_n}{dt} + \frac{1}{2}\left(u_{n+1}^2-u_{n-1}^2\right) = 0.
\end{equation}
Surprisingly, in addition to the regular rarefaction and dispersive shock, many other interesting dispersive wave structures such as the traveling dispersive shock wave have been numerically observed in \eqref{eq: first discretization scheme}. Moreover, an alternative discretization scheme of the Hopf equation has also been proposed and studied in \cite{LAX1986250},
\begin{equation}\label{eq: int dis intro}
    \frac{du_n}{dt} + u_n\left(u_{n+1}-u_{n-1}\right) = 0.
\end{equation}
The lattice in Eq.~\eqref{eq: int dis intro} has shown to be integrable by performing a variable transformation \cite{KAC1975160} to a model which is directly relevant with the Toda lattice. Based on the numerical studies \cite{https://doi.org/10.1111/sapm.12767} of the Riemann problem associated with this alternative discretization \eqref{eq: int dis intro}, it shows the emergence of both the rarefaction and dispersive shock. However, this paper shall focus on a non-integrable variant of the lattice \eqref{eq: int dis intro}.

This paper is structured as follows. In section \ref{sec: model des}, we discuss, in detail, the discrete lattice model which will be the focus of this paper, and meanwhile propose two quasi-continuum models in section \ref{sec: qua-cont appro}, which will be utilized to approximate the DSW and RW of the discrete lattice. We then attempt to analytically derive, in section \ref{sec: tw solutions}-\ref{sec: periodic waves}, the traveling and periodic waves of both quasi-continuum models. Next, in section \ref{sec: conservation laws}, we list some important conservation laws of the two quasi-continuum models, which serve as the prerequisites to derive the system of Whitham modulation equations. In section \ref{sec: Whitham modulation equations}, we perform the detailed analysis and derivation of the modulation equations which describe the slow spatial-temporal evolution of all the relevant parameters of the periodic waves. The DSW fitting is done in section \ref{sec: DSW fitting} so that we shall obtain some analytical predictions on multiple edge features of the DSW, and they are to be compared with those numerically estimated DSW edge features that we shall perform in the later section \ref{sec: numerical vali}. We next conduct some necessary numerical comparisons in the rest of the sections. Namely, in section \ref{sec: RWs}, we first compute the analytical self-similar solutions of the quasi-continuum models, and then compare such self-similar solutions with the rarefaction waves of the discrete lattice model and also the two quasi-continuum models. On the other hand, in section \ref{sec: numerical vali}, we discuss some important methods of performing numerical estimation of different edge features of the dispersive shock wave, and then we compare these numerically estimated DSW edge features between the two quasi-continuum models and the discrete lattice. Finally, we end this paper in section \ref{sec: conclu and future direc} with some comments about this work and also propose some potential future directions.

\section{Model description}\label{sec: model des}

\begin{figure}[b!]
    \centering
    \includegraphics[width=0.4\linewidth]{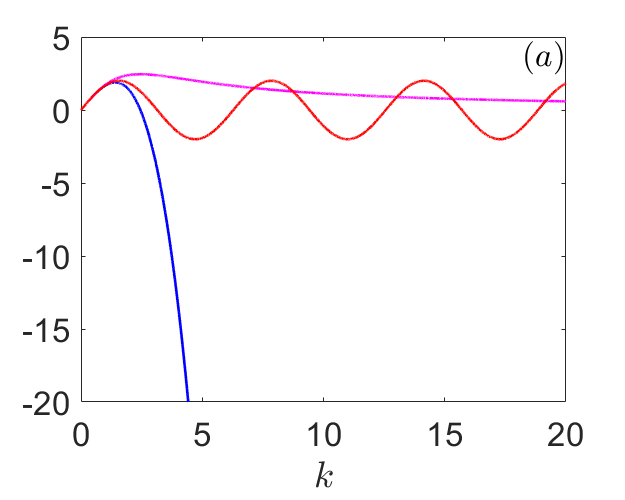}
    \includegraphics[width=0.4\linewidth]
    {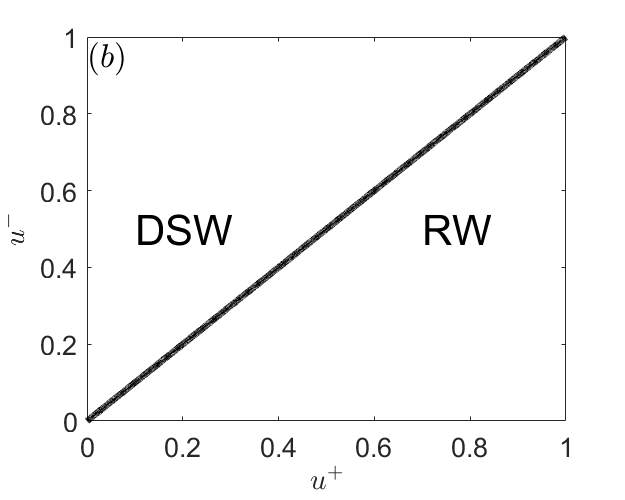}
    \caption{(a) The linear dispersion relation curves of the lattice \eqref{eq: extension 2} and two quasi-continuum models Eqs.~\eqref{eq: Non-regularized model}, \eqref{eq: BBM model}. (b) the classification of wave structures in the lattice \eqref{eq: extension 2}.}
    \label{fig:Wave classification}
\end{figure}

In this paper, we focus on a non-integrable variant of the integrable lattice in Eq.~\eqref{eq: int dis intro},
which is a discretization of the following continuum Hopf equation,
\begin{equation}\label{eq: Standard Hopf equation}
    u_t + \left(u^2\right)_x = 0.
\end{equation}
Before we display our lattice model, it is worthwhile to note that one can extend the scaler conservation law in Eq.~\eqref{eq: Standard Hopf equation} as follows,
\begin{equation}\label{eq: general discrete conservation law}
    u_t + \left[\Phi(u)\right]_x = 0,
\end{equation}
where $\Phi$ denotes a continuously differentiable potential function. With the chain rule, the discrete and dispersive regularization of the model Eq.~\eqref{eq: general discrete conservation law} reads,
\begin{equation}
    \frac{du_n}{dt} + \frac{1}{2}\Phi'(u_n)\left(u_{n+1}-u_{n-1}\right) = 0.
\end{equation}
To further demonstrate that the lattice in Eq.~\eqref{eq: int dis intro} is dispersive, we look for a plane-wave solution in the form of $u_n(t) = \overline{u} + ae^{i\left(kn-\omega t\right)}$, where $0 < a \ll 1$ is a small parameter, a direct substitution of this ansatz into \eqref{eq: int dis intro} yields the following linear dispersion relation,
\begin{equation}\label{eq: ldr for inter dis model}
    \omega\left(k,\overline{u}\right) = 2\overline{u}\sin(k),
\end{equation}
so clearly the linear dispersion relation in Eq.~\eqref{eq: ldr for inter dis model} does not always have zero second derivative with respect to $k$, and hence this shows that the lattice \eqref{eq: int dis intro} is dispersion.

One can readily extend the integrable system Eq.~\eqref{eq: int dis intro} as follows,
\begin{equation}\label{eq: extension 2}
    \frac{d u_{n}}{dt} + \left(u_n\right)^{\widetilde{p}}\left(u_{n+1} - u_{n-1}\right) = 0.
\end{equation}
where $\widetilde{p}$ is an arbitrary positive constant.

We notice first that Eq.~\eqref{eq: extension 2} is a discretization of the following continuum conservation law,
\begin{equation}
    u_t + \frac{2}{\widetilde{p}+1}\left(u^{\widetilde{p}+1}\right)_x = 0.
\end{equation}
Similarly, looking for a plane-wave ansatz yields the following linear dispersion relation for the lattice \eqref{eq: extension 2},
\begin{equation}\label{eq: linear dr for discrete lattice}
    \omega_0\left(\overline{u},k\right) = 2\overline{u}^{\widetilde{p}}\sin\left(k\right).
\end{equation}

We note that integrability may no longer persist in system \eqref{eq: extension 2} when $\widetilde{p} \neq 1$. However, relevant numerical studies have confirmed the emergence of both DSW and RW. As a consequence, it is necessary to find some quasi-continuum models in order to quasi-analytically capture and describe the characteristics of both the DSW and RW observed in system \eqref{eq: extension 2}. In this paper, we are interested in the so-called Riemann problem associated with the discrete lattice \eqref{eq: extension 2}. That is, we solve the initial value problem of Eq.~\eqref{eq: extension 2} with the following step-like initial data,
\begin{equation}\label{eq: Riemann initial data}
    u_n(0) = \begin{cases}
        u^-, \quad n \leq 0,\\
        u^+, \quad n > 0.
    \end{cases}
\end{equation}
As we shall see, when we have the upward step ($u^- > u^+$), a DSW shall form, while a RW emerges if we instead have the downward step ($u^- < u^+$).

\begin{figure}[b!]
    \centering
    \includegraphics[width=0.99\linewidth]{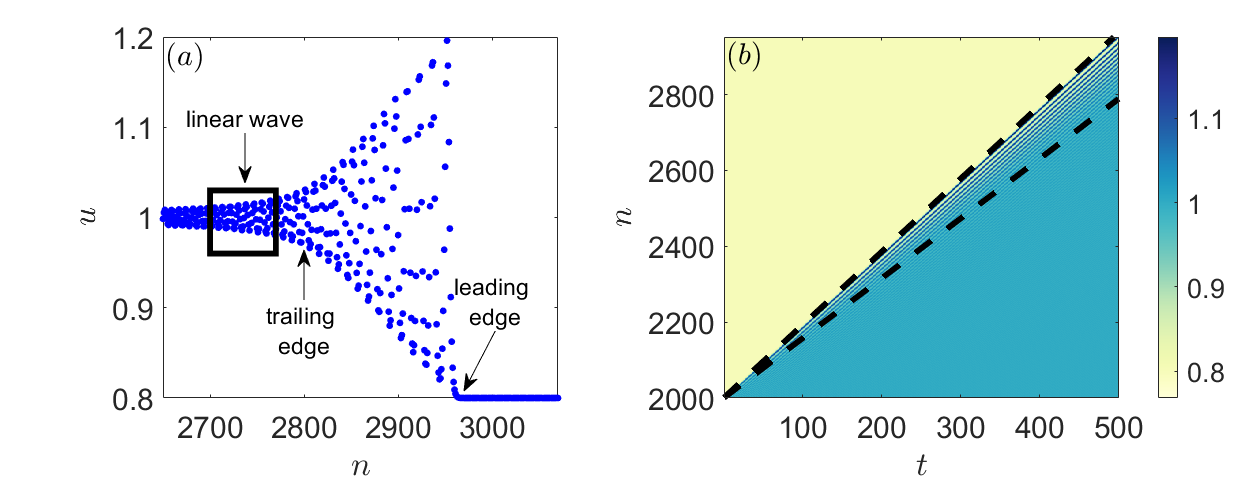}
    \caption{Dispersive shock wave of the lattice \eqref{eq: extension 2}. The left panel (a) depicts the spatial profile of the DSW of the lattice \eqref{eq: extension 2} at $t = 500$ for $\widetilde{p} = 1/2$, while the right panel (b) showcases the space-time plot of the associated DSW. Notice that the two dashed black lines in the panel (b) represent the estimation of the leading and the trailing edges of the DSW based on the DSW fitting theoretical predictions from section \ref{sec: DSW fitting}. }
    \label{fig:Lattice DSW profile versus density plot}
\end{figure}

\section{Quasi-continuum approximations}\label{sec: qua-cont appro}
To find some relevant quasi-continuum approximations for the system \eqref{eq: extension 2}, we take the following two slow variables on space and time,
\begin{equation}\label{eq: Slow variables}
X = \epsilon n, \quad T = \epsilon t,
\end{equation}
where $0 < \epsilon \ll 1$ is a formal smallness parameter.
Because of the definitions of $X,T$ in Eq.~\eqref{eq: Slow variables}, we then know
\begin{equation}\label{eq: Relations, wavenumber and freq}
    K = \epsilon^{-1}k, \quad \Omega = \epsilon^{-1}\omega.
\end{equation}

Then, we can rewrite Eq.~\eqref{eq: extension 2} as follows,
\begin{equation}\label{eq:ansatz substitution}
    \epsilon u_{T} + u^{\widetilde{p}}\left(u\left(X+\epsilon\right) - u\left(X-\epsilon\right)\right) = 0.
\end{equation}
We Taylor expand Eq.~\eqref{eq:ansatz substitution} and collect terms up to order of $\mathcal{O}(\epsilon^3)$ to obtain the following \textit{non-regularized} continuum model,
\begin{equation}\label{eq: Non-regularized model}
    u_T + u^{\widetilde{p}}\left(2u_X + \frac{\epsilon^2}{3}u_{XXX}\right) = 0.
\end{equation}
Eq.~\eqref{eq: Non-regularized model} is the first quasi-continuum model that we shall focus throughout this work. We then derive the linear dispersion relation of Eq.~\eqref{eq: Non-regularized model} by taking the plane-wave ansatz $u(X,T) = A + Be^{i\left(KX-\Omega T\right)}$, where $0 < B \ll 1$ is a small parameter. Substitution of the plane-wave ansatz into Eq.~\eqref{eq: Non-regularized model} yields the following linear dispersion relation,
\begin{equation}\label{eq: Linear dispersion relation of non-reg model}
    \Omega\left(A,K\right) = 2A^{\widetilde{p}}K\left(1 - \frac{\epsilon^2K^2}{6}\right).
\end{equation}

Moreover, we can regularize the model \eqref{eq: Non-regularized model} via reducing the order of its spatial derivatives. To this end, we rewrite Eq.~\eqref{eq: Non-regularized model} as follows,
\begin{equation}\label{eq: KdV-like model}
    \left(u^{1-\widetilde{p}}\right)_{T} = -2\left(1-\widetilde{p}\right)\left(1+\frac{\epsilon^{2}}{6}\partial^{2}_{X}\right)u_{X}.
\end{equation}
Inverting the operator $1 + \frac{\epsilon^{2}}{6}\partial_X^{2}$ on the right hand side of Eq.~\eqref{eq: KdV-like model} yields the following \textit{regularized} quasi-continuum model,
\begin{equation}\label{eq: BBM model}
    \left(u^{1-\widetilde{p}}\right)_{T} - \frac{\epsilon^{2}}{6}\left(u^{1-\widetilde{p}}\right)_{XXT} = -2\left(1-\widetilde{p}\right)u_X.
\end{equation}
Then, we compute the corresponding linear dispersion relation of the regularized by substituting the plane-wave ansatz, which then gives,
\begin{equation}\label{eq: Linear dispersion rela of the regularized model}
    \Omega(A,K) = \frac{2A^{\widetilde{p}}K}{1 + \frac{\epsilon^{2}K^2}{6}}.
\end{equation}

The figure \ref{fig:linear dr} depicts the linear dispersion relations of the three models Eqs.~\eqref{eq: extension 2}, \eqref{eq: Non-regularized model} and \eqref{eq: BBM model}. We can clearly see that the dispersion curve (blue) of the non-regularized model \eqref{eq: Non-regularized model} asymptotes to negative infinity drastically as the wavenumber $k$ increases, while that of the regularized model \eqref{eq: BBM model} always remains finite with a much smaller magnitude. Hence, we confirm the importance of performing regularization on the non-regularized model \eqref{eq: Non-regularized model}.

\section{Traveling-wave solutions}\label{sec: tw solutions}

For both quasi-continuum models of Eqs.~\eqref{eq: Non-regularized model} and \eqref{eq: BBM model}, we first look for traveling solitary waves in the form of 
\begin{equation}\label{eq: traveling sw}
    u\left(X,T\right) = U\left(Z\right), \quad Z = X - cT,
\end{equation}
where $c$ denotes the speed of propagation of the traveling wave.

\paragraph{Non-regularized model.} For the non-regularized model \eqref{eq: Non-regularized model}, substituting the traveling-wave ansatz \eqref{eq: traveling sw} into Eq.~\eqref{eq: Non-regularized model} yields,
\begin{equation}\label{eq: Non-reg cotraveling ODE}
    -cU^{-\widetilde{p}}U_Z + 2U_Z + \frac{\epsilon^2}{3}U_{ZZZ} = 0.
\end{equation}
A direct integration of Eq.~\eqref{eq: Non-reg cotraveling ODE}, with the requests for all constants of integration being zero, yields the following periodic solutions,
\begin{equation}\label{eq: Non-reg solitary soln}
    U(Z) = \bigg(\frac{c}{(1-\widetilde{p})(2-\widetilde{p})}\bigg)^{\frac{1}{\widetilde{p}}}\cos^{\frac{2}{\widetilde{p}}}\left(\frac{\sqrt{3}\widetilde{p}}{\sqrt{2}\epsilon}Z\right).
\end{equation}
Clearly, we notice that the solution in Eq.~\eqref{eq: Non-reg solitary soln} do not hold for $\widetilde{p} = 1,2$ as it becomes singular. Moreover, the periodic solution in Eq.~\eqref{eq: Non-reg solitary soln} can be used to describe the traveling solitary waves by only limiting the domain of the solution to one specific period.

\paragraph{Regularized model.} For the regularized model \eqref{eq: BBM model}, we first apply the change of variable $v = u^{1-\widetilde{p}}$, and then substitute the ansatz in terms of $V(Z) = U(Z)^{1-\widetilde{p}}$ into Eq.~\eqref{eq: BBM model} to obtain that, upon integration with respect $Z$ twice,
\begin{equation}\label{eq:BBM co-traveling ODE}
    \frac{\epsilon^2c}{6}\left(V_Z\right)^2 = cV^2 - \frac{4\left(1-\widetilde{p}\right)^2}{\left(2-\widetilde{p}\right)}V^{\frac{2-\widetilde{p}}{1-\widetilde{p}}} + 2AV + B,
\end{equation}
where $A,B$ are two constants of integration.

With the relation that $U = V^{\frac{1}{1-\widetilde{p}}}$, we set $A = B = 0$, and then a direct integration of Eq.~\eqref{eq:BBM co-traveling ODE} yields,
\begin{equation}\label{eq: analytical solitary wave for BBM model}
    U(Z) = \left(\frac{c\left(2-\widetilde{p}\right)}{4\left(1-\widetilde{p}\right)^2}\right)^{\frac{1}{\widetilde{p}}}\text{sech}^{\frac{2}{\widetilde{p}}}\left(\frac{\sqrt{6}\widetilde{p}}{2\left(1-\widetilde{p}\right)\epsilon}Z\right).
\end{equation}
Clearly, we notice that Eq.~\eqref{eq: analytical solitary wave for BBM model} does not hold when $\widetilde{p} = 1$, but we are not going to consider the case of $\widetilde{p} = 1$ as it associates with the integrable lattice \eqref{eq: int dis intro}.

Finally, it is important and worthwhile to notice that the traveling-wave solutions in Eqs.~\eqref{eq: Non-reg solitary soln} and \eqref{eq: analytical solitary wave for BBM model} for the two respective quasi-continuum models can be complex for some cases of $\widetilde{p}$, but in this paper we are only interested in the scenarios where we have real wave field of $U$ and hence ignore those cases of $\widetilde{p}$ which lead to a complex wave. Meanwhile, since the goal of this paper is to study the nonlinear dispersive waves of the discrete lattice \eqref{eq: extension 2} through the two quasi-continuum models, we will not perform any numerical studies of the traveling solitary waves in \eqref{eq: extension 2} nor its comparison to the two solitary waves we found in Eqs.~\eqref{eq: Non-reg solitary soln} and \eqref{eq: analytical solitary wave for BBM model}.

\section{Periodic solutions}\label{sec: periodic waves}

In this section, we shall derive the periodic traveling wave solutions for the two quasi-continuum models. Periodic traveling wave solutions are prerequisite for Whitham analysis of the DSW as we shall see later in the section \ref{sec: Whitham modulation equations}.

\begin{figure}[t!]
    \centering
    \includegraphics[width=0.4\linewidth]{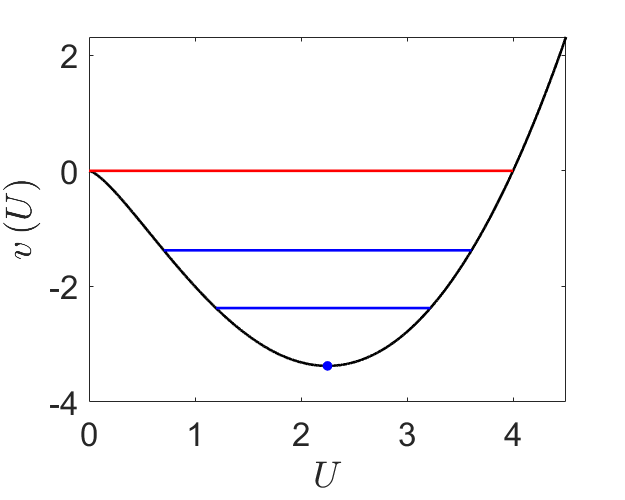}
    \caption{The potential curve of the non-regularized model \eqref{eq: Non-regularized model}. }
    \label{fig:Potential curve of the non-reg model}
\end{figure}

\paragraph{Non-regularized model.} For the non-regularized model \eqref{eq: Non-regularized model}, unfortunately the periodic solutions are not analytically obtainable, to the best of our knowledge. However, we are still capable to show the existence of the periodic waves in the non-regularized model \eqref{eq: Non-regularized model}. To this end, we display the potential curve of the model and address some relevant periodic orbits which serve as the representatives of the periodic wave solutions. Figure \ref{fig:Potential curve of the non-reg model} depicts the potential curve of the non-regularized model, which is,
\begin{equation}
    v(U) =  \frac{-2c}{\left(1-\widetilde{p}\right)\left(2-\widetilde{p}\right)}U^{2-\widetilde{p}}+2U^2-2AU-B,
\end{equation}
where $A = B = 0$ and $c = 3/2$ in figure \ref{fig:Potential curve of the non-reg model}. We notice that the red line showcases the homoclinic orbit which refers to the traveling solitary wave, while the two blue lines denote two periodic orbits that indicate the associated periodic solutions in the non-regularized model \eqref{eq: Non-regularized model}.

\paragraph{Regularized model.} For the regularized model \eqref{eq: BBM model}, such the solutions are only analytically derivable for two specific cases of $\widetilde{p}$: $\widetilde{p} = 1/2, 2/3$. We divide the details of the derivation of the periodic solutions into the following two subsections.

\paragraph{{\Large $\widetilde{p} = 1/2$}.} When $\widetilde{p} = 1/2$, we can rewrite the ODE \eqref{eq:BBM co-traveling ODE} as follows,
\begin{equation}\label{eq: case of p_t = 1}
    \begin{aligned}
    \left(V_Z\right)^2 &= \frac{4}{\epsilon^2c}\left(-V^3+\frac{3c}{2}V^2+3AV+\frac{3B}{2}\right),\\
    &= \frac{4}{\epsilon^2c}\left(V_1-V\right)\left(V_2-V\right)\left(V_3-V\right),
    \end{aligned}
\end{equation}
where $V_1 \le V_2 \le V_3$ denote the three roots of the polynomial of $P\left(V\right) = - V^{3} + \frac{3c}{2}V^{2} + 3AV + \frac{3B}{2}$.

A direct integration of the ODE \eqref{eq: case of p_t = 1}
 yields,
 \begin{equation}\label{eq: periodic solution for p_t = 1}
     V\left(Z\right) = V_2 + \left(V_3 - V_2\right)\text{cn}^{2}\left(\frac{\sqrt{V_3 - V_1}}{\epsilon\sqrt{c}}\left(Z-Z_0\right), m\right)
 \end{equation}
where $m = \frac{V_3 - V_2}{V_3 - V_1}$ denotes the elliptic modulus, and $Z_0$ is a constant of integration.

Since $U^{1/2} = V$, the periodic solution in terms of $U$ reads, 
\begin{equation}\label{eq: periodic wave for p = 1/2}
    U(Z) = \left[V_2+\left(V_3-V_2\right)\text{cn}^{2}\left(\frac{\sqrt{V_3-V_1}}{\epsilon \sqrt{c}}\left(Z-Z_0\right),m\right)\right]^2.
\end{equation}

Furthermore, we notice that one can infer the associated soliton amplitude of the periodic solution of $U$ in Eq.~\eqref{eq: periodic wave for p = 1/2}. To this end, note that at the solitonic limit $m \to 1$, the periodic wave for $V$ in Eq.~\eqref{eq: periodic solution for p_t = 1} simply reduces to the following solitary wave,
\begin{equation}
    V(Z) = V_2 + (V_3 - V_2)\text{sech}^2\left(\frac{\sqrt{V_3-V_1}}{\epsilon\sqrt{c}}Z\right),
\end{equation}
where we have ignored the arbitrary initial phase parameter $Z_0$, and therefore the soliton amplitude, denoted by $a^{+}$, is given as follows,
\begin{equation}\label{eq: soliton amplitude for p_t = 1}
    a^{+} = V_3 - V_2.
\end{equation}

To find an explicit expression of the amplitude $a^{+}$ in terms of the soliton speed $c$, we notice that by expanding the product of $\left(V_1 - V\right)\left(V_2 - V\right)\left(V_3 - V\right)$ and equating the associated coefficients with that of the polynomial $P\left(V\right)$ defined above, 
\begin{align}
    V_1 + V_2 + V_3 = \frac{3c}{2}.
\end{align}
Since at the solitonic limit, $m \to 1$, or equivalently $U_1 \to U_2$, we know the final explicit soliton amplitude formula for $V$ is given as follows,
\begin{equation}\label{eq: explicit formula for a+ for p_t = 1}
    a^{+}_V = \frac{3c}{2} - 3V_2.
\end{equation}
On the other hand, since $U = V^2$, we know the corresponding soliton-amplitude relation for $U$ reads,
\begin{equation}
    a^+_U = \frac{1}{2}\left[\left(\frac{3c}{2}-2V_2\right)^2 - V_2^2\right].
\end{equation}

\paragraph{{\Large$\widetilde{p} = 2/3$}. } On the other hand, for the case of $\widetilde{p} = 2/3$, we can rewrite the ODE \eqref{eq:BBM co-traveling ODE} as follows,
\begin{equation}\label{eq: co-traveling ODE for p_t = 2}
    \begin{aligned}
    \left(V_Z\right)^2 &= -\frac{2}{\epsilon^2 c}\left(V^{4}-3cV^2-6AV-3B\right),\\
    &= -\frac{2}{\epsilon^2 c}\left(V-V_1\right)\left(V-V_2\right)\left(V-V_3\right)\left(V-V_4\right),
    \end{aligned}
\end{equation}
where $V_1 \le V_2 \le V_3 \le V_4$ refer to the four roots of the polynomial $P\left(V\right) = V^{4} - 3cV^{2} - 6AV - 3B$.

Before we integrate the Eq.~\eqref{eq: co-traveling ODE for p_t = 2}, we denote 
\begin{equation}\label{eq: dispersion term def}
    \mu = -\frac{2}{\epsilon^{2}c}.
\end{equation}

If $\mu > 0$, then the oscillatory behavior of the periodic wave occurs in the interval $V_2 \le V \le V_3$, and then a direct integration of the ODE \eqref{eq: co-traveling ODE for p_t = 2} yields the following solution in terms of Jacobi elliptic function,
\begin{equation}\label{eq: solution ass. with posi dispersion}
    V\left(Z\right) = V_2 + \frac{\left(V_3 - V_2\right)\text{cn}^{2}\left(\zeta, m\right)}{1 - \frac{V_3 - V_2}{V_4 - V_2}\text{sn}^{2}\left(\zeta, m\right)}, 
\end{equation}
where 
\begin{equation}\label{eq: Defs of phase and elliptic modulus}
\begin{aligned}
    \zeta &= \frac{\sqrt{\left|\mu\right|\left(V_3 - V_1\right)\left(V_4 - V_2\right)}Z}{2},\\
    m &= \frac{\left(V_3 - V_2\right)\left(V_4 - V_1\right)}{\left(V_4 - V_2\right)\left(V_3 - V_1\right)}.
\end{aligned}
\end{equation}
We observe that one can readily deduce the soliton amplitude by taking advantage of the periodic solution given in Eq.~\eqref{eq: solution ass. with posi dispersion}. To this end, we notice that as the solitonic limit, one must have that either $V_2 \to V_1$ or $V_3 \to V_4$.

In the former case of $V_2 \to V_1$, we end up with the following bright solitary wave,
\begin{equation}
    V(Z) = V_1 + \frac{V_3 - V_1}{\text{cosh}^{2}\left(\zeta\right) - \frac{V_3 - V_1}{V_4 - V_1}\text{sinh}^{2}\left(\zeta\right)}.
\end{equation}
On the other hand, for the latter case of $U_3 \to U_4$, we arrive at the following dark solitary wave solution,
\begin{equation}
    V(Z) = V_4 - \frac{V_4 - V_2}{\text{cosh}^{2}\left(\zeta\right) - \frac{V_4 - V_2}{V_4 - V_1}\text{sinh}^{2}\left(\zeta\right)},
\end{equation}
We next focus on the case when $\mu < 0$. We make the observation that the oscillation then occurs in the interval $V_3 \le V \le V_4$. A direct integration of the ODE \eqref{eq: co-traveling ODE for p_t = 2} yields,
\begin{equation}\label{eq: periodic solution ass. negative dispersion}
    V(Z) = V_3 + \frac{\left(V_4 - V_3\right)\text{cn}^{2}\left(\zeta, m\right)}{1 + \frac{V_4 - V_3}{V_3 - V_1}\text{sn}^{2}\left(\zeta, m\right)},
\end{equation}
where $\zeta$ is given in Eq.\eqref{eq: Defs of phase and elliptic modulus} and the elliptic modulus $m$ now becomes
\begin{equation}\label{eq: ellip modu ass. neg dispersion}
    m = \frac{\left(V_4 - V_3\right)\left(V_2 - V_1\right)}{\left(V_4 - V_2\right)\left(V_3 -V_1\right)}.
\end{equation}
Then, we observe that since $U = V^3$, we then know the periodic solution to $U$ reads,
\begin{equation}
    U(Z) = \left[V_3 + \frac{\left(V_4-V_3\right)\text{cn}^2\left(\zeta,m\right)}{1+\frac{V_4-V_3}{V_3-V_1}\text{sn}^{2}\left(\zeta,m\right)}\right]^3.
\end{equation}

Now, to derive the associated soliton amplitude, we observe that at the solitonic limit, $m \to 1$, or equivalently $V_3 \to V_2$, and the associated periodic wave in Eq.~\eqref{eq: periodic solution ass. negative dispersion} reduces to 
\begin{equation}
    V(Z) = V_2 + \frac{V_4 - V_2}{\text{cosh}^{2}\left(\zeta\right) + \frac{V_4 - V_2}{V_2 - V_1}\text{sinh}^{2}\left(\zeta\right)}.
\end{equation}
Hence, the amplitude $a^{+}$ reads,
\begin{equation}\label{eq: soliton amplitude}
    a^{+} = V_4 - V_2.
\end{equation}
To obtain the the speed-amplitude relation between $a^+$ and $c$, we notice that by expanding the product of $\left(V - V_1\right)\left(V - V_2\right)\left(V - V_3\right)\left(V - V_4\right)$ and equating the coefficients with that of $P\left(V\right)$, we have that
\begin{equation}\label{eq: parameters relation}
   \begin{aligned}
    V_1V_2 + V_1V_3 + V_2V_3 + V_1V_4 + V_2V_4 + V_3V_4 &= -3c,\\
    V_1 + V_2 + V_3 + V_4 &= 0.
    \end{aligned}
\end{equation}
With the relations in Eqs.~\eqref{eq: parameters relation} and $V_3 = V_2$, at the solitonic limit, we can infer that the soliton amplitude for the periodic solution of $V$ is given as follows,
\begin{equation}\label{eq: soliton amplitude in explicit form}
    a_V^{+} = \sqrt{3c - 2\left(V_2\right)^{2}} - 2V_2.   
\end{equation}
On the other hand, since $U = V^3$, for the solitonic amplitude for the periodic solution to $U$, it is accordingly given as follows,
\begin{equation}\label{eq: soliton amplitude for U when p = 2/3}
    a_U^+ = \frac{1}{2}\left[\left(\sqrt{3c-2\left(V_2\right)^2} - V_2\right)^3 - \left(V_2\right)^3\right].
\end{equation}

\section{Conservation laws}\label{sec: conservation laws}

In this section, we display some conservation laws of the two quasi-continuum models. Conservation laws are crucially important in the theoretical analysis of the DSW as the so-called Whitham modulation equations shall be derived from these necessary conservation laws of the model. We will illustrate the important role that Whitham modulation equations play in the analysis of the DSW in section \ref{sec: Whitham modulation equations}. 
\paragraph{Non-regularized model.} Firstly, for the non-regularized model \eqref{eq: Non-regularized model}, it possesses the following two conservation laws,
\begin{equation}\label{eq: Conservation laws of non-reg model}
    \begin{aligned}
    \frac{1}{1-\widetilde{p}}\left(u^{1-\widetilde{p}}\right)_{T}  +2u_{X} + \frac{\epsilon^{2}}{3}u_{XXX} &= 0,\\
    \frac{1}{2-\widetilde{p}}\left(u^{2-\widetilde{p}}\right)_{T} +\left(u^{2}\right)_X + \frac{\epsilon^{2}}{3}\left(uu_{XX}\right)_{X} - \frac{\epsilon^{2}}{6}\left[\left(u_X\right)^{2}\right]_{X} &= 0.
    \end{aligned}
\end{equation}
We note that the first equation in system \eqref{eq: Conservation laws of non-reg model} refers to the conservation of \textit{mass}, while the second equation is the conservation of \textit{momentum}.

However, it is important to notice that the system of conservation laws in \eqref{eq: Conservation laws of non-reg model} do not hold when $\widetilde{p} = 1,2$. Instead, for these two particular cases, the two associated conservation laws read: For $\widetilde{p} = 1$,
\begin{equation}\label{eq: p = 1 non-reg conservations}
    \begin{aligned}
        \left(\log(u)\right)_T + 2u_X + \frac{\epsilon^2}{3}u_{XXX} &= 0,\\
        u_T + \left(u^2\right)_X + \frac{\epsilon^2}{3}\left(uu_{XX}\right)_X - \frac{\epsilon^2}{6}\left[\left(u_X\right)^2\right]_X &= 0.
    \end{aligned}
\end{equation}
For $\widetilde{p} = 2$, 
\begin{equation}\label{eq: p = 2 non-reg conservations}
    \begin{aligned}
       -\left(\frac{1}{u}\right)_T + 2u_X + \frac{\epsilon^2}{3}u_{XXX} &= 0,\\
       \left(\log(u)\right)_T + \left(u^2\right)_X + \frac{\epsilon^2}{3}\left(uu_{XX}\right)_X - \frac{\epsilon^2}{6}\left[\left(u_X\right)^2\right]_X &= 0.
    \end{aligned}
\end{equation}

\paragraph{Regularized model.} On the other hand, for the regularized model \eqref{eq: BBM model}, we have the following conservation laws,
\begin{align}\label{eq: Conservation law of BBM model}
    \begin{aligned}
    \left[u^{1-\widetilde{p}} - \frac{\epsilon^2}{6}\left(u^{1-\widetilde{p}}\right)_{XX}\right]_T + 2\left(1-\widetilde{p}\right)u_X &= 0,\\
    \left[\frac{1}{2}\left(u^{1-\widetilde{p}}\right)^2 + \frac{\epsilon^2}{12}\left[\left(u^{1-\widetilde{p}}\right)_X\right]^2\right]_T + \left[\frac{2\left(1-\widetilde{p}\right)}{\left(2-\widetilde{p}\right)}u^{2-\widetilde{p}}-\frac{\epsilon^2}{6}u^{1-\widetilde{p}}\left(u^{1-\widetilde{p}}\right)_{XT}\right]_X &= 0.
    \end{aligned}
\end{align}
Notice that the above system does not hold for $\widetilde{p} = 2$ due to the singularity in the second equation of \eqref{eq: Conservation law of BBM model}. Instead, the two conservation laws for the regularized model \eqref{eq: BBM model} for $\widetilde{p} = 2$ read,
\begin{equation}
    \begin{aligned}
        \left[u^{-1} - \frac{\epsilon^2}{6}\left(u^{-1}\right)_{XX}\right]_T - 2u_X &= 0,\\
        \left[\frac{1}{2}u^{-2} + \frac{\epsilon^2}{12}\left[\left(u^{-1}\right)_X\right]^2\right]_T + \left[-2\log(u)-\frac{\epsilon^2}{6}u^{-1}\left(u^{-1}\right)_{XT}\right]_X &= 0.
    \end{aligned}
\end{equation}
Similarly, the first and the second equation in system \eqref{eq: Conservation law of BBM model} refer to the conservation of mass and momentum, respectively.

\section{Whitham modulation equations}\label{sec: Whitham modulation equations}

In this section, we perform Whitham analysis on the periodic wave solutions of the quasi-continuum model Eqs.~\eqref{eq: Non-regularized model} and \eqref{eq: BBM model} as this is an indispensable process before we can analytically capture the behaviors of the DSWs. We first notice that the periodic wave of the model Eq.~\eqref{eq: BBM model} is a three-parameters family solution as the co-traveling frame ODE Eq.~\eqref{eq:BBM co-traveling ODE} contains three parameters of $c, A, B$ where we recall that $c$ denotes the speed of propagation of the wave and $A,B$ are two constants of integration. This suggests that to capture the complete dynamics of these parameters of the periodic wave, we need three modulation equations to form a complete closed system.

To derive the Whitham modulation equations, we notice that the two spatial and temporal variables $X,T$ are already slow variables, and then we introduce the following fast phase variable $\theta$,
\begin{equation}
    \theta = \frac{KX - \Omega T}{\epsilon},
\end{equation}
and we seek for periodic wave solution in the following multiple-scale form of
\begin{equation}
    u(X,T) = \phi\left(\theta;X,T\right) + \epsilon u_1\left(\theta; X,T\right) + \mathcal{O}\left(\epsilon^2\right),
\end{equation}
where $\phi\left(\theta;X,T\right)$ refers to a periodic traveling wave with a fixed $2\pi$ period and we recall that $0 < \epsilon \ll 1$ is the same smallness parameter that we applied in the previous sections. 

Next, the fast phase variable $\theta$ is further defined by $\theta_X = K / \epsilon$ and $\theta_T = -\Omega / \epsilon$, where both now the wavenumber $K$ and the frequency $\Omega$ are functions of both slowly varying variables of $X, T$. We then observe that the compatible condition of $\theta_{XT} = \theta_{TX}$ yields the following equation,
\begin{equation}\label{eq: conservation of waves}
    K_{T} + \Omega_{X} = 0,
\end{equation}
which is the so-called \textit{conservation of waves}. It is the first modulation equation in our closed modulation system. 

To derive the remaining modulation equations, we apply the method of averaging the conservation laws \cite{EL201611}. To this end, we define the following averaging operation: For any function $F$,
\begin{equation}\label{eq: Averaging operation}
    \overline{F\left(\phi\right)} = \frac{1}{2\pi}\int_{0}^{2\pi}F\left(\phi(\theta)\right)d\theta.
\end{equation}
We then apply this averaging operation to the conservation laws we derived in section \ref{sec: conservation laws} for both quasi-continuum models and then collect at the leading order to obtain the remanining modulation equations.

\paragraph{Modulation system of the non-regularized model.} On the one hand, when $\widetilde{p} \neq 1,2$, the modulation system for the non-regularized model \eqref{eq: Non-regularized model} reads, by dropping all the higher order terms in $\epsilon$,
\begin{equation}\label{eq: Modulation system for non-reg model}
\begin{aligned}
    K_{T} + \Omega_{X} &= 0,\\
    \frac{1}{1-\widetilde{p}}\left(\overline{\phi^{1-\widetilde{p}}}\right)_{T} + 2\left(\overline{\phi}\right)_X &= 0,\\
    \frac{1}{2-\widetilde{p}}\left(\overline{\phi^{2-\widetilde{p}}}\right)_{T} + \left(\overline{\phi^{2}} - \frac{K^2}{2}\overline{\phi_{\theta}^{2}}\right)_{X} &= 0.
\end{aligned}
\end{equation}
In addition, when $\widetilde{p} = 1,2$, the two corresponding modulation systems read,
\begin{equation}\label{eq: Modulation system for p = 1, non-reg model}
    \begin{aligned}
    K_T + \Omega_X &= 0,\\
    \left(\overline{\log(\phi)}\right)_T + 2\left(\overline{\phi}\right)_X &= 0,\\
    \overline{\phi}_T + \left(\overline{\phi^2} - \frac{K^2}{2}\overline{\left(\phi_\theta\right)^2}\right)_X &= 0,
    \end{aligned}
\end{equation}
and
\begin{equation}\label{eq: Modulation system for p = 2, non-reg model}
    \begin{aligned}
       K_T + \Omega_X &= 0,\\
        -\left(\overline{\phi^{-1}}\right)_T + 2\left(\overline{\phi}\right)_X &= 0,\\
        \left(\overline{\log(\phi)}\right)_T + \left(\overline{\phi^2} - \frac{K^2}{2}\overline{\left(\phi_\theta\right)^2}\right)_X &= 0,
    \end{aligned}
\end{equation}
respectively.

\paragraph{Modulation system of the regularized model.} On the other hand, the modulation system for the regularized model \eqref{eq: BBM model} reads, by ignoring all the higher order terms in $\epsilon$, when $\widetilde{p} \neq 2$,
\begin{equation}\label{eq: Whitham system for BBM model}
\begin{aligned}
    K_T + \Omega_X &= 0,\\
    \left(\overline{\phi^{1-\widetilde{p}}}\right)_T + 2\left(1-\widetilde{p}\right)\left(\overline{\phi}\right)_X &= 0,\\
    \left[\frac{1}{2}\overline{\phi^{2\left(1-\widetilde{p}\right)}} + \frac{K^2}{12}\left[\left(\overline{\phi^{1-\widetilde{p}}}\right)_\theta\right]^2\right]_T + \left[\frac{2\left(1-\widetilde{p}\right)}{\left(2-\widetilde{p}\right)}\overline{\phi^{2-\widetilde{p}}} - \frac{K\Omega}{6}\overline{\left[\left(\phi^{1-\widetilde{p}}\right)_\theta\right]^2}\right]_X &= 0. 
\end{aligned}
\end{equation}
When $\widetilde{p} = 2$, the modulation system reads,
\begin{equation}
    \begin{aligned}
    K_T + \Omega_X &= 0,\\
    \left(\overline{\phi^{-1}}\right)_T -2\left(\overline{\phi}\right)_X &= 0,\\
    \bigg(\frac{1}{2}\overline{\phi^{-2}}+\frac{K^2}{12}\overline{\left[\left(\phi^{-1}\right)_\theta\right]^2}\bigg)_T + \bigg(-2\overline{\log\left(\phi\right)} - \frac{K\Omega}{6}\overline{\left[\left(\phi^{-1}\right)_\theta\right]^2}\bigg)_X &= 0.
    \end{aligned}
\end{equation}

\section{DSW fitting}\label{sec: DSW fitting}

In this section, we apply the so-called DSW fitting method \cite{EL201611} to obtain some useful analytical insights on some specific properties of the dispersive shock waves. The properties include the edge speeds and the edge wavenumber of the DSW. We later shall compare these analytical foundings with the numerical results to verify the performance of this theoretical method.

\subsection{Discrete lattice model}
We first perform the DSW fitting \cite{EL201611} on the discrete lattice \eqref{eq: extension 2}.
We notice that the edge information of the DSW of the discrete lattice model can be obtained upon solving the following two initial value problems which are the so-called \textit{simple wave} ODEs \cite{EL201611,10.1063/1.1947120},
\begin{equation}\label{eq: DSW fitting ODEs}
    \begin{aligned}
        \frac{dk}{d\overline{u}} &= \frac{\partial_{\overline{u}}\omega_0}{2\overline{u}^{\widetilde{p}} - \partial_k\omega_0}, \hspace{5mm} k(u^{+}) = 0,\\
        \frac{d\widetilde{k}}{d\overline{u}} &= \frac{\partial_{\overline{u}}\widetilde{\omega}_s}{2\overline{u}^{\widetilde{p}} - \partial_{\widetilde{k}}\widetilde{\omega}_s}, \hspace{5mm} \widetilde{k}(u^{-}) = 0,
    \end{aligned}
\end{equation}
where $\omega_0$ refers to the linear dispersion relation in Eq.~\eqref{eq: linear dr for discrete lattice} of the lattice and $\widetilde{\omega}_s$ denotes the following conjugate dispersion relation of the lattice \eqref{eq: extension 2},
\begin{equation}\label{eq: conjugate dr}
    \widetilde{\omega}_s\left(\overline{u}, \widetilde{k}\right) = -i \omega_0\left(\overline{u}, i\widetilde{k}\right)=2\overline{u}^{\widetilde{p}}\text{sinh}(\widetilde{k}).
\end{equation}
A direct integration of the two boundary value problems in Eqs.~\eqref{eq: DSW fitting ODEs} yields,
\begin{equation}\label{eq: Solutions to the BVPs} 
  \begin{aligned}
  k\left(\overline{u}\right) &= 2\cos^{-1}\left(\left(\frac{\overline{u}}{u^{+}}\right)^{-\frac{\widetilde{p}}{2}}\right),\\
  \widetilde{k}\left(\overline{u}\right) &= 2\text{cosh}^{-1}\left(\left(\frac{\overline{u}}{u^{-}}\right)^{-\frac{\widetilde{p}}{2}}\right).
  \end{aligned}
\end{equation}
Moreover, the edge-speed predictions of the lattice DSW are given as follows,
\begin{equation}\label{eq: edge speeds}
    \begin{aligned}
        s^{-} &= \frac{\partial\omega_0}{\partial k}\left(k^{-}, u^{-}\right) = 4\left(u^{+}\right)^{\widetilde{p}} - 2\left(u^{-}\right)^{\widetilde{p}},\\
        s^{+} &= \frac{\widetilde{\omega}_s}{\widetilde{k}}\left(\widetilde{k}^{+}, u^{+}\right)=\frac{4\left(u^{+}u^{-}\right)^{\frac{\widetilde{p}}{2}}\sqrt{\left(\frac{u^{-}}{u^{+}}\right)^{\widetilde{p}} - 1}}{\widetilde{k}^{+}}.
    \end{aligned}
\end{equation}
where $\widetilde{k}^{+} = \widetilde{k}\left(u^{+}\right)$.

\subsection{Non-regularized model}

Similarly, to perform DSW fitting for the non-regularized model \eqref{eq: Non-regularized model}, we solve initial-value problems in system \eqref{eq: DSW fitting ODEs}, with the lattice linear dispersion relation $\omega_0\left(\overline{u},k\right)$ replaced with $\Omega_0\left(\overline{u},K\right)$ in Eq.~\eqref{eq: Linear dispersion relation of non-reg model}.
We notice further that the conjugate dispersion relation for the non-regularized model \eqref{eq: Non-regularized model} reads,
\begin{equation}\label{eq: Conjugate dispersion rela for non-reg model}
\Omega_s(\overline{u},\widetilde{K}) = -i\Omega_0(\overline{u}, i\widetilde{K}) = 2\overline{u}^{\widetilde{p}}\widetilde{K}\left(1 + \frac{\epsilon^{2}\widetilde{K}^2}{6}\right).
\end{equation}
It then follows that,
\begin{equation}
    \begin{aligned}
        &\left(1 - \frac{\epsilon^2K^2}{6}\right)^{-\frac{3}{2\widetilde{p}}} = \frac{\overline{u}}{u^+},\\
        &\left(1 + \frac{\epsilon^2\widetilde{K}^2}{6}\right)^{-\frac{3}{2\widetilde{p}}} = \frac{\overline{u}}{u^-}.
    \end{aligned}
\end{equation}
Since $K^- = K(u^-)$ and $\widetilde{K}^+ = \widetilde{K}(u^+)$, we know
\begin{equation}
    \begin{aligned}
        &\left(K^-\right)^2 = \frac{6}{\epsilon^2}\left(1 - \left(\frac{u^+}{u^-}\right)^{\frac{2\widetilde{p}}{3}}\right),\\
        &\left(\widetilde{K}^+\right)^2 = \frac{6}{\epsilon^2}\left(\left(\frac{u^-}{u^+}\right)^{\frac{2\widetilde{p}}{3}}-1\right),
    \end{aligned}
\end{equation}
and the edge speeds are given as follows,
\begin{equation}
    \begin{aligned}
    &s^{-} = \partial_K\Omega_0\left(u^-,K^-\right) = 2\left(u^-\right)^{\widetilde{p}}\left(3\left(\frac{u^+}{u^-}\right)^{\frac{2\widetilde{p}}{3}}-2\right),\\
    &s^+ = \frac{\Omega_s}{\widetilde{K}}\left(u^+,\widetilde{K}^+\right) = 2\left(u^+\right)^{\widetilde{p}}\left(\frac{u^-}{u^+}\right)^{\frac{2\widetilde{p}}{3}}.
    \end{aligned}
\end{equation}

\subsection{Regularized model Eq.~\eqref{eq: BBM model}}

Finally, for the DSW fitting of the continuum regularized model \eqref{eq: BBM model}, we again solve the two initial-value problems in system \eqref{eq: DSW fitting ODEs} with the linear dispersion relation given by $\Omega_0(\overline{u},K)$ in Eq.~\eqref{eq: Linear dispersion rela of the regularized model}. Meanwhile, notice that the associated conjugate dispersion of the regularized model \eqref{eq: BBM model} is given as follows,
\begin{equation}
    \widetilde{\Omega}_s\left(\overline{u},\widetilde{K}\right) = -i\Omega_0\left(\overline{u}, i\widetilde{K}\right) = \frac{2\overline{u}^{\widetilde{p}}\widetilde{K}}{1 - \frac{\epsilon^{2}\widetilde{K}^{2}}{6}}.
\end{equation}
Similarly, this yields,
\begin{equation}
    \begin{aligned}
        \log\left|\overline{u}\right| &= \frac{1}{\widetilde{p}}\left(\frac{\epsilon^{2}K^{2}}{12} + \log\left(6 + \epsilon^{2}K^{2}\right)\right) + C,\\
        \log\left|\overline{u}\right| &= \frac{1}{\widetilde{p}}\left(-\frac{\epsilon^{2}\widetilde{K}^{2}}{12} + \log\left(6 - \epsilon^{2}\widetilde{K}^{2}\right)\right) + \widetilde{C},
    \end{aligned}
\end{equation}
where $C, \widetilde{C}$ are two constants of integration.

Applying the initial conditions, we end up with,
\begin{equation}\label{eq: solutions to BVPs}
    \begin{aligned}
        \log\left|\frac{\overline{u}}{u^{+}}\right| &= \frac{1}{\widetilde{p}}\left(\frac{\epsilon^{2}K^{2}}{12} + \log\left(1 + \frac{\epsilon^{2}K^{2}}{6}\right)\right),\\
        \log\left|\frac{\overline{u}}{u^{-}}\right| &= \frac{1}{\widetilde{p}}\left(-\frac{\epsilon^{2}\widetilde{K}^{2}}{12} + \log\left(1 - \frac{\epsilon^{2}\widetilde{K}^{2}}{6}\right)\right).
    \end{aligned}
\end{equation}

Then, we know at the two edges of the DSW, the wavenumbers are given respectively as, $K^{-} = K\left(u^{-}\right)$, and $\widetilde{K}^{+} = \widetilde{K}\left(u^{+}\right)$, and the associated trailing and leading edge speeds, denoted by $s^{-}, s^{+}$ respectively, are given as follows,
\begin{equation}\label{eq: edge speeds}
    \begin{aligned}
        s^{-} &= \frac{\partial\Omega_0}{\partial K}\left(u^{-}, K^{-}\right) = \frac{2\left(u^{-}\right)^{\widetilde{p}}\left(1 - \frac{\epsilon^{2}\left(K^{-}\right)^{2}}{6}\right)}{\left(1 + \frac{\epsilon^{2}\left(K^{-}\right)^{2}}{6}\right)^{2}},\\
        s^{+} &= \frac{\widetilde{\Omega}_s\left(u^{+}, \widetilde{K}^{+}\right)}{\widetilde{K}^{+}} = \frac{2\left(u^{+}\right)^{\widetilde{p}}}{1 - \frac{\epsilon^{2}\left(\widetilde{K}^{+}\right)^{2}}{6}}.
    \end{aligned}
\end{equation}

Finally, we recall that for the regularized model \eqref{eq: BBM model}, when $\widetilde{p} = 1,2$, we also have the theoretical predictions on the DSW solitonic amplitude based on the speed-amplitude relations derived in Eqs.~\eqref{eq: explicit formula for a+ for p_t = 1} and \eqref{eq: soliton amplitude in explicit form}. Namely, when $\widetilde{p} = 1/2,2/3$, the theoretical predictions on the DSW amplitude read,
\begin{equation}\label{eq: DSW-fitting for solitonic amplitude}
    \begin{aligned}
    &a^+_{\widetilde{p} = 1/2} = \frac{1}{2}\left[\left(\frac{3s^+}{2}-2\sqrt{u^+}\right)^2 - u^+\right], \\
    &a^+_{\widetilde{p} = 2/3} = \frac{1}{2}\left[\left(\sqrt{3s^+-2\left(u^+\right)^{2/3}} - \left(u^+\right)^{1/3}\right)^3 - u^+\right].
    \end{aligned}
\end{equation}

\section{Rarefaction waves}\label{sec: RWs}

\begin{figure}[b!]
    \centering
    \includegraphics[width=0.4\linewidth]{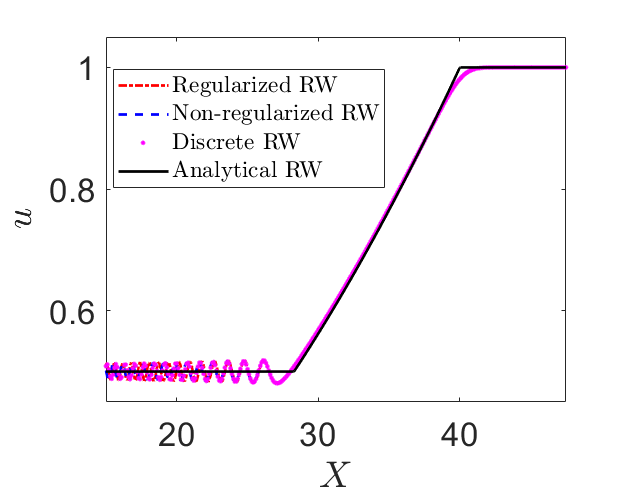}
    \caption{Comparison of the rarefaction waves: on the one hand, the solid black curve refers to the analytical self-similar solution in Eq.~\eqref{eq: Rarefactions soln}. On the other hand, the red dotted-dashed, blue dashed, and magenta dots depict the numerical rarefaction waves associated with the regularized model \eqref{eq: BBM model}, non-regularized model \eqref{eq: Non-regularized model}, and the discrete lattice \eqref{eq: extension 2}, respectively. Moreoever, we have the two parameters $\epsilon = 0.1, \widetilde{p} = 1/2$, and the dynamics is shown at $T = 20 (t = 200)$.}
    \label{fig:RW comparison}
\end{figure}

In this section, we discuss in detail, both analytically and numerically, the rarefaction waves emerging from the numerical simulation. We notice that the rarefaction wave can be described by the self-similar solutions of the dispersionless system of the PDE which, for both Eqs.~\eqref{eq: Non-regularized model} and \eqref{eq: BBM model}, reads
\begin{equation}\label{eq: Dispersionless system}
    u_T + 2u^{\tilde{p}}u_X = 0,
\end{equation}
with the upward step-like initial condition given by 
\begin{equation}\label{eq: Upward IC}
    u(X,0) = 
    \begin{cases}
        u^{-}, \hspace{5mm}X \leq 0,\\
        u^{+}, \hspace{5mm}X > 0,
    \end{cases}
\end{equation}
where now $u^{-} < u^{+}$.

To compute the self-similar solutions to Eq.~\eqref{eq: Dispersionless system}, we assume the following self-similar ansatz,
\begin{equation}\label{eq: Self-similar ansatz}
    u(X,T) = S\left(\kappa\right), \quad \kappa = \frac{X}{T}.
\end{equation}
Substitution of the ansatz \eqref{eq: Self-similar ansatz} into the Eq.~\eqref{eq: Dispersionless system} yields
\begin{equation}\label{eq: self-similar ODE}
    -\kappa S_\kappa + 2 S^{\tilde{p}}S_\kappa = 0.
\end{equation}
We then solve for $S$ in Eq.~\eqref{eq: self-similar ODE} to obtain the following analytical self-similar solution,
\begin{equation}\label{eq: Rarefactions soln}
    u(X,T) = 
    \begin{cases}
        u^{-}, \quad X \leq 2T\left(u^{-}\right)^{\widetilde{p}},\\
        \left(\frac{X}{2T}\right)^{\frac{1}{\tilde{p}}}, \quad 2T\left(u^{-}\right)^{\widetilde{p}} < X \leq 2T\left(u^{+}\right)^{\widetilde{p}},\\
        u^{+},\quad X > 2T\left(u^{+}\right)^{\widetilde{p}}.
    \end{cases}
\end{equation}
Figure \ref{fig:RW comparison} shows the comparison of the numerical rarefaction waves with the analytical self-similar solution in Eq.~\eqref{eq: Rarefactions soln} for the case of $\widetilde{p} = 1/2$. From the figure \ref{fig:RW comparison}, we can clearly see the close alignment of the analytical solution with the its numerical correspondences. This suggests that the two quasi-continuum models provide a reasonable approximation to the rarefaction waves observed numerically in the discrete lattice \eqref{eq: extension 2}.

\section{Numerical validation}\label{sec: numerical vali}

\begin{figure}[b!]
    \centering
    \includegraphics[width=0.4\linewidth]{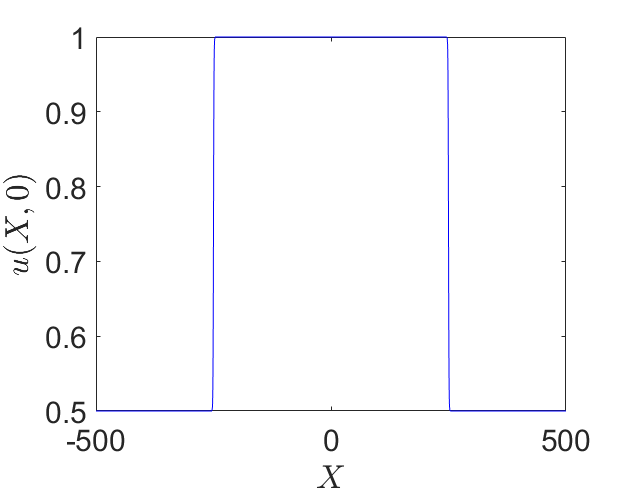}
    \caption{The box-type initial condition in Eq.~\eqref{eq: Box-type IC}. Notice that the two backgrounds are given as $u^- = 1, u^+ = 0.5$ and the two location parameters are $a = -250, b = 250$. Finally, $\delta = 1$.}
    \label{fig:Riemann box-type initial data}
\end{figure}

In this section, we perform numerical simulations of the Riemann problems associated with both quasi-continuum models of Eqs.~\eqref{eq: Non-regularized model}, \eqref{eq: BBM model} and the discrete lattice \eqref{eq: extension 2}. Namely, we numerically solve Eq.~\eqref{eq: BBM model} with the following box-type initial condition,
\begin{equation}\label{eq: Box-type IC}
    u\left(X,0\right) = u^{+} - \left(\frac{u^{+} - u^{-}}{2}\right)\bigg[\tanh\bigg(\delta\left(X - a\right)\bigg) - \tanh\bigg(\delta\left(X - b\right)\bigg)\bigg],
\end{equation}
where $a<b$ are two real numbers representing the left and right positions of the box, and $\delta > 0$ is parameter that serves as smoothing the jumps between the states of $u^-, u^+$. Figure \ref{fig:Riemann box-type initial data} depicts the spatial profile of the initial data in Eq.~\eqref{eq: Box-type IC}, and it is essentially a combination of two initial conditions: One with the upward step which shall lead to the emergence of the rarefaction, and another with a upward initial step which will lead to the evolution of the DSW. We notice that we will apply the RK$4$ time integrating scheme with the pseudo-spectral discretization of the space to numerically solve both quasi-continuum. Since the pseudo-spectral discretization method assumes the periodic boundary conditions, this illustrate our choice of the box-type initial data in Eq.~\eqref{eq: Box-type IC} for both quasi-continuum models. On the other hand, for the discrete lattice \eqref{eq: BBM model}, we utilize the RK$4$ time integration method to numerically solve it. In addition, notice that we utilize $\epsilon = 0.1$ in all the relevant numerical simulations throughout this work.

\subsection{Numerical preliminaries}

Before we present all the numerical results, we briefly mention the methods which we applied to numerically compute the three edge features of the DSW including: The leading and trailing edge speeds, denoted as $s^+$ and $s^-$, respectively, and the trailing-edge wavenumber $K^-$.

\paragraph{Leading-edge speed.} For $s^+$, we first pinpoint the leading-edge location. We treat the associated $x$ coordinate of the highest peak of the DSW as the location of the leading edge, and we denote this specific point as $X^+$. Then the speed $s^+$ is simply calculated as follows,
\begin{equation}\label{eq: leading-edge speed}
    s^+ = \frac{X^+ - b}{t_{f}},
\end{equation}
where $t_f$ refers to the total simulation time. Meanwhile, it is important to notice that when computing $s^+$ of the DSW of the lattice model \eqref{eq: extension 2}, we have to further divide the $s^+$ in Eq.~\eqref{eq: leading-edge speed} by $\epsilon$, as $T = \epsilon t$, and according to the chain rule, $\frac{dX}{dT} = \epsilon^{-1}\frac{dX}{dt}$.

\paragraph{Trailing-edge speed.} For $s^-$, we simply need to replace the $X^+$ by $X^-$, which is the trailing-edge location of the DSW. However, it is more complex to measure the trailing-edge location $X^-$ and its measurement cannot be exact. Alternative there shall be various elaborate methods to estimate $X^-$, we provide a relatively easy way to estimate $X^-$. Firstly, we define the following two quantities,
\begin{equation}\label{eq: trailing-edge measure quantities}
   \begin{aligned}
    &I^u = u^- + \frac{|u^- - u^+|}{N},\\
    &I^l = u^- - \frac{|u^- - u^+|}{N},\\
    \end{aligned}
\end{equation}
where $N$ is a positive integer. Typically, we utilize $N = 8$ in our numerical simulations. Then, we extrapolate the local maxima and minima of the DSW within the windows of $\left(I^u + \delta,I^u - \delta\right)$, $\left(I^l - \delta, I^l + \delta\right)$, where $\delta$ is some small numbers which can be choosen to be $\delta = \frac{|u^- - u^+|}{10}$. Next, we apply a least-square based method to fit two straight lines through these local maxima and the local minima so that these two lines shall intersect at a point in the vicinity of the trailing edge of the DSW. We treat such the point of intersection as the trailing-edge location of the DSW and we denote it by $X^-$. To compute the trailing-edge speed $s^-$ of the DSW, we use the same formula as in Eq.~\eqref{eq: leading-edge speed} with $X^+$ replaced with $X^-$ to measure the trailing-edge speed $s^-$.

\begin{figure}[t!]
    \centering
    \includegraphics[width=0.8\linewidth]{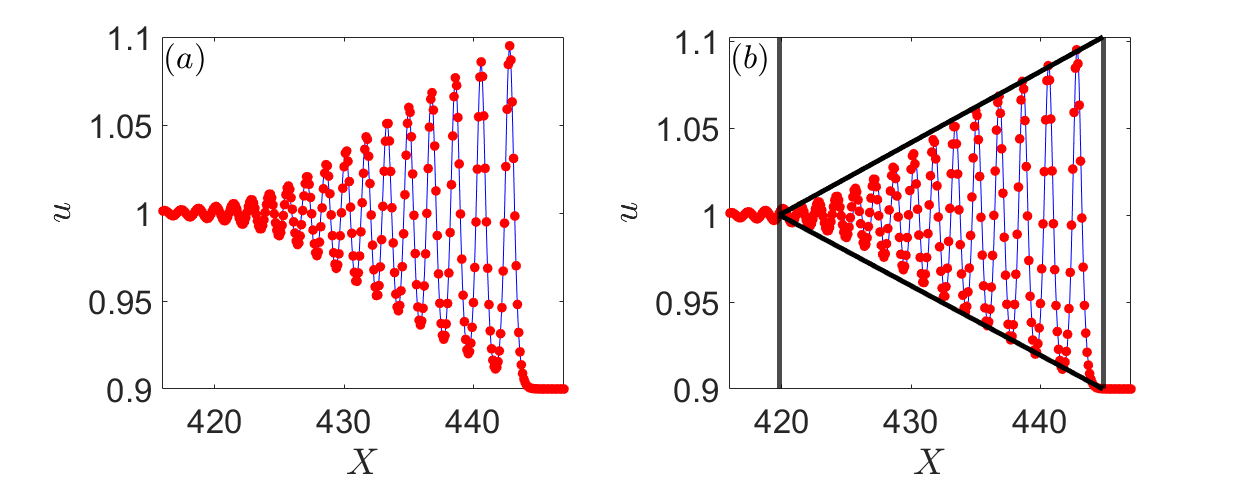}
    \hfill
    \includegraphics[width=0.8\linewidth]{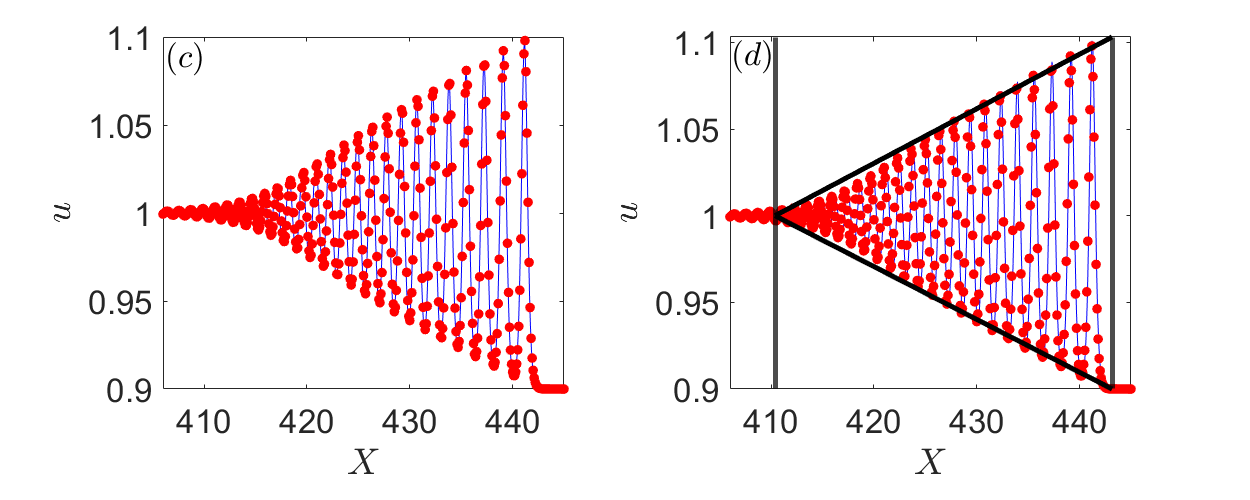}
    \caption{DSW spatial comparison at $T = 150 (t = 1500)$. Notice that the first column depicts the comparison of the DSW of the non-regularized model \eqref{eq: Non-regularized model} with that of the lattice \eqref{eq: extension 2}, while the second column displays the comparison of the DSW between the regularized model \eqref{eq: BBM model} and the lattice \eqref{eq: extension 2}. Meanwhile, panels (a)-(b) correspond to the case of $\widetilde{p} = 1/2$, while panels (c)-(d) associate with the case of $\widetilde{p} = 2/3$, where all the dynamics are shown at $T= 150 (t = 1500)$. Finally, note that the red dots refer to the lattice DSW data, while the solid blue curve denote that of the quasi-continuum models.}
    \label{fig: DSW profile comparisons}
\end{figure}

\paragraph{Trailing-edge wavenumber.} For the trailing-edge wavenumber, we shall not apply the same way to measure such the quantity in both quasi-continuum models Eqs.~\eqref{eq: Non-regularized model}, \eqref{eq: BBM model} and the discrete lattice \eqref{eq: extension 2}. Firstly, for the quasi-continuum models, to estimate the trailing-edge wavenumber $K^-$, we first find the nearest local peak to the trailing-edge location $N^-$, and compute the associated $x$ coordinate of such the peak, denoted as $X_1^-$. Then, we find another local peak of the DSW which is just right next to $X_1^-$, and we denote its location as $X_2^-$. With these two locations of the adjacent peak, we can compute the wavelength $\lambda^- = X_2^- - X_1^-$ at the trailing edge, and then use the following formula to estimate $K^-$,
\begin{equation}\label{eq: Quasi-continuum models wavenumber estimation}
    K^- = \frac{2\pi}{\lambda^-}.
\end{equation}
On the other hand, for the DSW of the discrete lattice \eqref{eq: extension 2}, we use an alternative, but equivalent method. First we recall that our field variable is dependent on the co-traveling frame. Namely, $u_n(t) = u(kn - \omega t)$. Then, clearly,
\begin{equation}\label{eq: factoring out the frequency}
    u(kn-\omega t) = u\left(\omega\left(\frac{k}{\omega}n - t\right)\right).
\end{equation}
The key thought here is to measure the wave frequency $\omega$. This can be done in a similar fashion as we performed for the computation of the wavenumber of the qausi-continuum models in Eq.~\eqref{eq: Quasi-continuum models wavenumber estimation}. That is to say, we fix the spatial variable $n = n^-$, where $n^-$ refers to the trailing-edge location of the lattice DSW. Then, we observe that the field $u$ in Eq.~\eqref{eq: factoring out the frequency} is essentially a time-series data that only depend on $t$. Then we seek for the wavelength of the adjacent peaks of this time-series data, denoted as $\lambda^-_t = t^-_2 - t^-_1$. Then the wave frequency simply is estimated as $\omega = 2\pi / \lambda^-_t$. Once the wave frequency is known, the remaining task is to compute the quantity of $k/\omega$, since the multiplication of $k/\omega$ and the known $\omega$ will certainly yield the wavenumber $k$. To this end, we pick the grid $n_2^-$ which is just right next to the trailing-edge location $n^-$ so that $n_2^- = n^- + 1$. Then, we observe that the time-series data of $u_{n_2^-}(t)$ differs from that of $u_{n^-}(t)$ via a phase shift by $k/\omega$ amount. This phase shift can be easily measured (e.g. by looking at the $x$ vertical distance between the same local max of both $u_{n^-}(t)$ and $u_{n_2^-}(t)$), and we finally obtain our estimation for the trailing-edge wavenumber $k^-$ for the lattice DSW. However, a nature question can be why we do not simply try to apply the formula in Eq.~\eqref{eq: Quasi-continuum models wavenumber estimation} to calculate $k^-$? The reason for this is that when we perform the simulation of the discrete lattice \eqref{eq: extension 2}, since $X = \epsilon n$, we know the lattice spacing is $\epsilon$. As mentioned before, $\epsilon = 0.1$ in all the relevant simulations, so this indicates that the spatial resolution may not be well enough for us to apply the method in Eq.~\eqref{eq: Quasi-continuum models wavenumber estimation} for wavenumber estimation of the lattice DSW. However, on the other hand, the temporal grid surely has much better resolution than the spatial grid does as the temporal grid spacing, which is the time step, is much smaller than the spatial lattice spacing in the numerical simulation of the lattice \eqref{eq: extension 2}, and the wavenumber estimated with this temporal frequency approach will yield much accurate results.

\subsection{Numerical comparisons}

\begin{figure}[t!]
    \centering
    \includegraphics[width=1.05\linewidth]{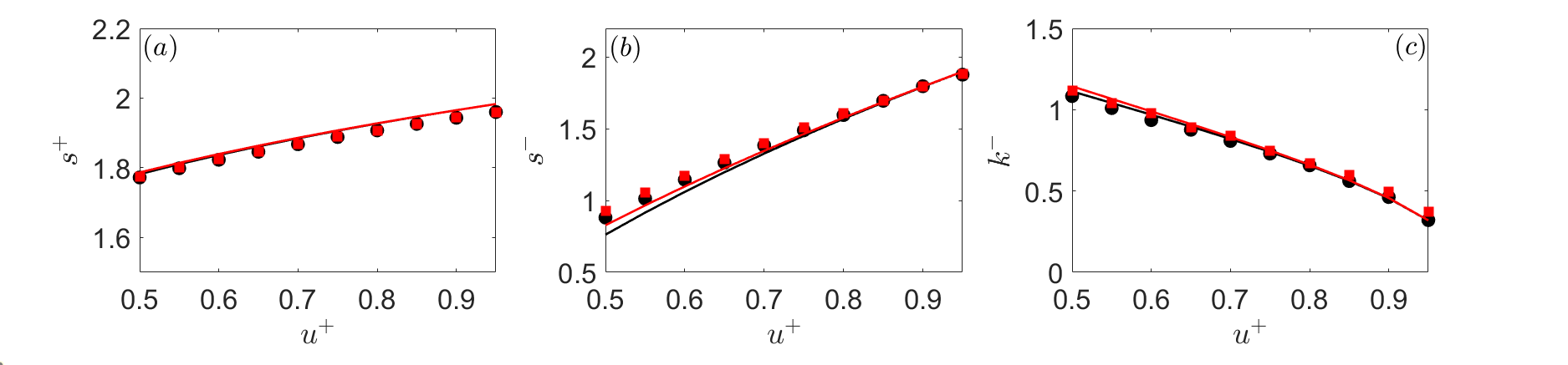}
    \hfill
    \includegraphics[width=1.05\linewidth]{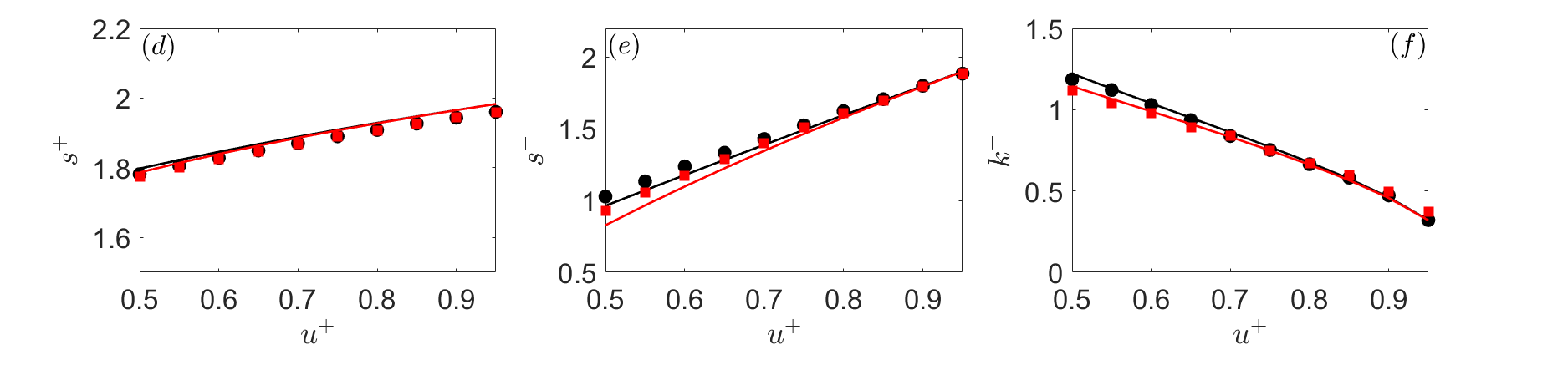}
    \caption{The comparison of the DSW edge features for the case of $\widetilde{p} = 1/2$. Panels (a)-(c) represent the comparison between the non-regularized model \eqref{eq: Non-regularized model} and the lattice \eqref{eq: extension 2}, while panels (d)-(f) display the comparison between the regularized model \eqref{eq: BBM model} and the lattice \eqref{eq: extension 2}. Notice that the solid red curves and the red dots in all the panels refer to the DSW fitting results on the lattice \eqref{eq: extension 2} and the numerical estimated edge features of the lattice DSW, while the solid black line and the black squares denote those of the quasi-continuum models.}
    \label{fig:comparison for p = 1/2}
\end{figure}

Firstly, we note that figure \ref{fig: DSW profile comparisons} depicts the comparison of the spatial profiles of the DSW between the quasi-continuum models \eqref{eq: Non-regularized model},\eqref{eq: BBM model} with that of the lattice \eqref{eq: extension 2}. For both cases of $\widetilde{p}$, we can see that the comparison is quite good as long as the jump, denoted as $\Delta = u^- - u^+$, is small. Furthermore, for both cases of $\widetilde{p} = 1/2,2/3$, in the comparison of the DSWs between the regularized model \eqref{eq: BBM model} and the lattice \eqref{eq: extension 2}, we have drawn a black triangular region which represents the spatial-profile prediction of the DSW based on the DSW-fitting theoretical results in section \ref{sec: DSW fitting}, and such the predicted spatial region of the DSW also fits well with the DSWs emerged from the simulation of both the quasi-continuum and lattice models.

\begin{figure}[t!]
    \centering
    \includegraphics[width=1.05\linewidth]{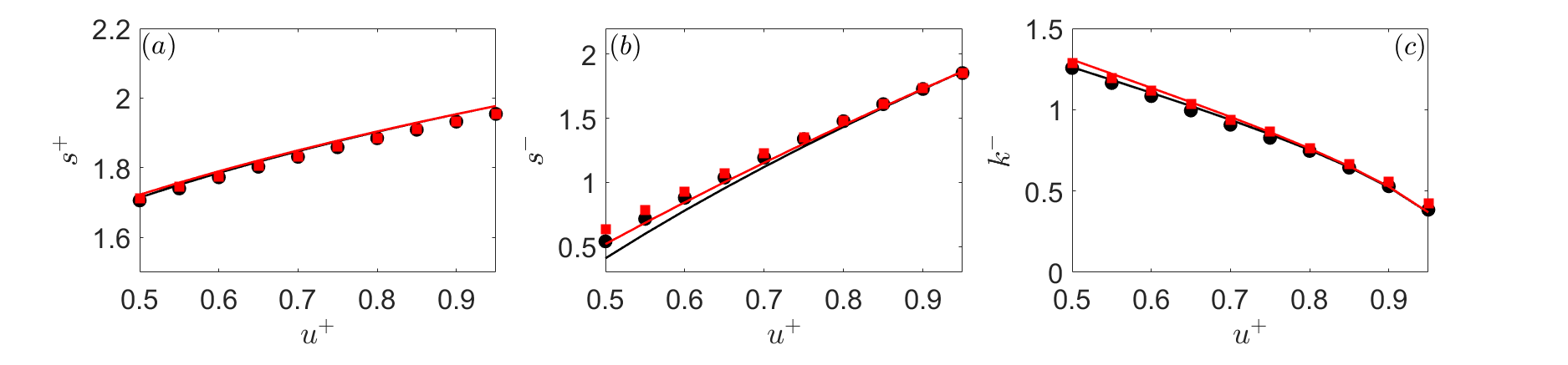}
    \hfill
    \includegraphics[width=1.05\linewidth]{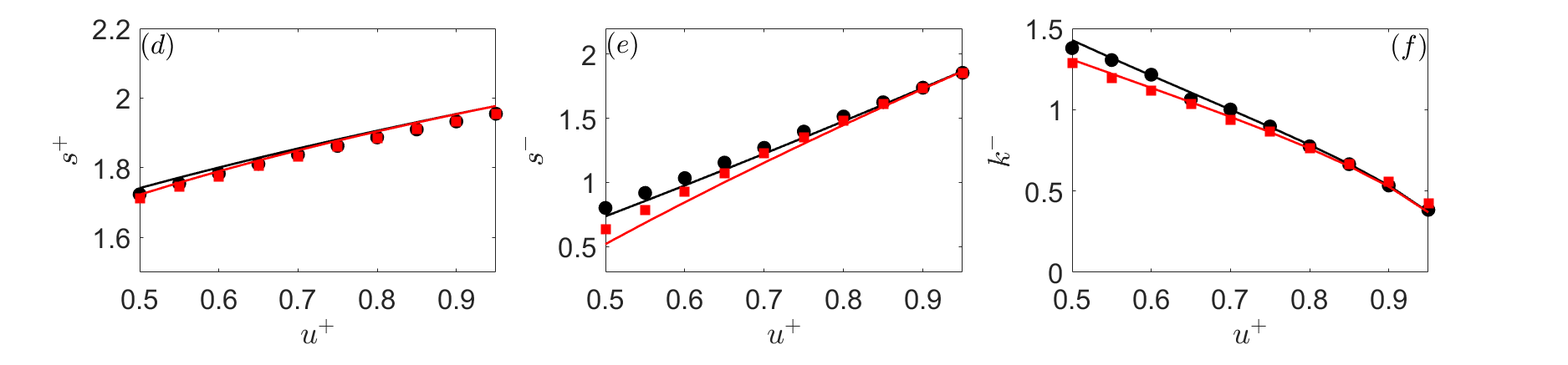}
    \caption{The comparison of the DSW edge features for the case of $\widetilde{p} = 2/3$. Panels (a)-(c) represent the comparison between the non-regularized model \eqref{eq: Non-regularized model} and the lattice \eqref{eq: extension 2}, while panels (d)-(f) display the comparison between the regularized model \eqref{eq: BBM model} and the lattice \eqref{eq: extension 2}. Notice that the solid red curves and the red dots in all the panels refer to the DSW fitting results on the lattice \eqref{eq: extension 2} and the numerical estimated edge features of the lattice DSW, while the solid black line and the black squares denote those of the quasi-continuum models.}
    \label{fig:comparison for p = 2}
\end{figure}

Furthermore, figures \ref{fig:comparison for p = 1/2} and \ref{fig:comparison for p = 2} showcase the comparison of numerically estimated DSW edge features with their associated DSW-fitting theoretical predictions based on section \ref{sec: DSW fitting}. We note that we have fixed the larger background $u^- = 1$ in all the relevant simulations and continuously varied the value of $u^+$ from $0.5$ to $0.95$ with a spacing being $0.05$. For the case of $\widetilde{p} = 1/2$, the comparisons are quite good for both quasi-continuum models with the lattice. Meanwhile, we have the expectation that as the jump $\Delta$ increases, the comparison shall become worse. However, for this particular case of $\widetilde{p}$, we can see that even for a relatively large jump $\Delta = 0.5$, the agreement of the lattice DSW and the quasi-continuum DSW edge features is still good. Also, notice that these discrete numerically measured DSW edge-feature data points closely aligned upon the solid lines which refer to their respective DSW-fitting theoretical predictions. This indicates that the DSW-fitting method indeed provides reasonable predictions on the DSW edge features. However, when $\widetilde{p} = 2/3$, we observe that the comparison is slightly less satisfactory than that of $\widetilde{p} = 1/2$. In particular, for example, for the comparison of the trailing-edge speed $s^-$ between the regularized model \eqref{eq: BBM model} and lattice \eqref{eq: extension 2}, when the jump $\Delta = 0.5$, the lattice DSW trailing-edge speed clearly deviates more from that of the regularized model DSW than the case of $\widetilde{p} = 1/2$. Meanwhile, we have also conducted the numerical comparison for the case of $\widetilde{p} = 2$ and we further found that the disagreement is even more prominent when the jump $\Delta = 0.5$. Nevertheless, we notice that the comparison is still good for both quasi-continuum models with the lattice when the jump $\Delta$ is small ($\Delta \sim 0.1$). Hence, these comparisons indicate that for $\widetilde{p} > 1$, both quasi-continuum models only provide a reasonably good approximation for the DSW of the lattice model as long as the jump parameter $\Delta$ is small.

\section{Conclusions and future directions}\label{sec: conclu and future direc}

In this paper, we revisit an integrable discrete lattice model which serves as a dispersive regulation of the dispersionless Hopf equation, and we have proposed and studied a non-integrable variant of such an integrable model. We conduct a systematic numerical studies on the evolution of the associated Riemann problems of the discrete lattice and discover only two dispersive wave structures shown in the phase diagram \ref{fig:Wave classification}. Then, we derive two quasi-continuum models to approximate these two wave structures in the lattice. Based on the two quasi-continuum models, we compute their respective traveling and periodic wave solutions for some particular cases of $\widetilde{p}$, provide some necessary and important conservation laws serving as the preliminaries of Whitham analysis. With the methodology of averaging the conservation laws \cite{whitham2011linear}, we derive the full closed modulation system for both quasi-continuum models and also perform reduction of the Whitham modulation equations near both the linear and solitonic limits to obtain the simple-wave ODEs which encode important information on the edge features of the DSW. We solve these simple-wave ODEs and gain theoretical predictions on multiple DSW edge features including the edge speeds and wavenumber. We conduct comparisons on not only the spatial profile of the DSW and RW between the lattice \eqref{eq: extension 2} and both the quasi-continuum models Eqs.~\eqref{eq: Non-regularized model}, \eqref{eq: BBM model}, but numerous different DSW edge features as well. We find that, through these numerical comparisons, both quasi-continuum models provide a reasonably well approximation to the DSW for situation where the values of parameter $\widetilde{p}$ are small, but only good approximation for $\widetilde{p} > 1$ if the jump $\Delta$ of the Riemann initial data is small.

There are still a variety of interesting open questions remaining, and we only list a few of them. Firstly, we have mentioned in the comparison section \ref{sec: numerical vali} that, for the case when $\widetilde{p} > 1$, both quasi-continuum models failed to yield a good approximation to the DSW of the lattice \eqref{eq: extension 2} if the jump $\Delta$ is large. This particular failure of the two quasi-continuum models will naturally lead one to think of other potential dispersive quasi-continuum models which can be possibly derived by performing a different set of change of spatial and temporal variables rather than Eq.~\eqref{eq: Slow variables} and also a possible scaling of the amplitude of the wave field of $u$. In addition, we notice that one can also investigate the lattice in Eq.~\eqref{eq: general discrete conservation law} where the potential $\Phi$ is not a power function. For instance, an interesting alternative potential can be the exponential function. The reason to change the potential $\Phi$ is as follows. We recall that we have only discovered the DSW and RW in the lattice \eqref{eq: general discrete conservation law} when a power-type potential is applied. Hence, a different type of potential may excite wave structures other than the DSW and RW so that the lattice becomes more versatile as a media to model various nonlinear dispersive waves. Lastly, it may also be relevant to consider some other discretization scheme for the scaler conservation law in Eq.~\eqref{eq: general discrete conservation law}. We notice that our lattice \eqref{eq: extension 2} uses the standard central difference formula to discretize the spatial derivative of $u_x$, so with a different numerical scheme, one shall obtain a distinct lattice, and it may have more abundant structures of wave. All these proposed and open directions are currently under consideration and will be reported in future publications.

\paragraph{Acknowledgment.} The author thanks Professor Panayotis G.~Kevrekidis from the University of Massachusetts Amherst for suggesting this interesting project.

\bibliography{main}

\begin{thebibliography}{10}

\bibitem{Hoefer:2009}
M.~Hoefer and M.~Ablowitz.
\newblock {D}ispersive shock waves.
\newblock {\em Scholarpedia}, 4(11):5562, 2009.
\newblock revision \#137922.

\bibitem{BIONDINI2024134315}
Gino Biondini, Christopher Chong, and Panayotis Kevrekidis.
\newblock On the whitham modulation equations for the toda lattice and the quantitative characterization of its dispersive shocks.
\newblock {\em Physica D: Nonlinear Phenomena}, 469:134315, 2024.

\bibitem{CHONG2024103352}
Christopher Chong, Ari Geisler, Panayotis~G. Kevrekidis, and Gino Biondini.
\newblock Integrable approximations of dispersive shock waves of the granular chain.
\newblock {\em Wave Motion}, 130:103352, 2024.

\bibitem{toda2012theory}
M.~Toda.
\newblock {\em Theory of Nonlinear Lattices}.
\newblock Springer Series in Solid-State Sciences. Springer Berlin Heidelberg, 2012.

\bibitem{Yang_Biondini_Chong_Kevrekidis_2025}
Su~Yang, Gino Biondini, Christopher Chong, and Panayotis~G. Kevrekidis.
\newblock A regularized continuum model for travelling waves and dispersive shocks of the granular chain.
\newblock {\em Journal of Nonlinear Waves}, 1:e2, 2025.

\bibitem{yang2025firstordercontinuummodelsnonlinear}
Su~Yang, Gino Biondini, Christopher Chong, and Panayotis~G. Kevrekidis.
\newblock First-order continuum models for nonlinear dispersive waves in the granular crystal lattice, 2025.

\bibitem{CHONG2022133533}
Christopher Chong, Michael Herrmann, and P.G. Kevrekidis.
\newblock Dispersive shock waves in lattices: A dimension reduction approach.
\newblock {\em Physica D: Nonlinear Phenomena}, 442:133533, 2022.

\bibitem{PhysRevE.95.062216}
H.~Yasuda, C.~Chong, J.~Yang, and P.~G. Kevrekidis.
\newblock Emergence of dispersive shocks and rarefaction waves in power-law contact models.
\newblock {\em Phys. Rev. E}, 95:062216, Jun 2017.

\bibitem{PhysRevLett.120.194101}
H.~Kim, E.~Kim, C.~Chong, P.~G. Kevrekidis, and J.~Yang.
\newblock Demonstration of dispersive rarefaction shocks in hollow elliptical cylinder chains.
\newblock {\em Phys. Rev. Lett.}, 120:194101, May 2018.

\bibitem{mohapatra2025dambreaksdiscretenonlinear}
Shrohan Mohapatra, Panayotis~G. Kevrekidis, Su~Yang, and Sathyanarayanan Chandramouli.
\newblock Dam breaks in the discrete nonlinear schr\"odinger equation, 2025.

\bibitem{PhysRevA.110.023304}
Sathyanarayanan Chandramouli, S.~I. Mistakidis, G.~C. Katsimiga, and P.~G. Kevrekidis.
\newblock Dispersive shock waves in a one-dimensional droplet-bearing environment.
\newblock {\em Phys. Rev. A}, 110:023304, Aug 2024.

\bibitem{Chandramouli_2023}
Sathyanarayanan Chandramouli, Nicholas Ossi, Ziad~H Musslimani, and Konstantinos~G Makris.
\newblock Dispersive hydrodynamics in non-hermitian nonlinear schrödinger equation with complex external potential.
\newblock {\em Nonlinearity}, 36(12):6798, nov 2023.

\bibitem{PhysRevLett.100.084504}
M.~A. Hoefer, M.~J. Ablowitz, and P.~Engels.
\newblock Piston dispersive shock wave problem.
\newblock {\em Phys. Rev. Lett.}, 100:084504, Feb 2008.

\bibitem{PhysRevE.75.021304}
E.~B. Herbold and V.~F. Nesterenko.
\newblock Shock wave structure in a strongly nonlinear lattice with viscous dissipation.
\newblock {\em Phys. Rev. E}, 75:021304, Feb 2007.

\bibitem{PhysRevE.80.056602}
Alain Molinari and Chiara Daraio.
\newblock Stationary shocks in periodic highly nonlinear granular chains.
\newblock {\em Phys. Rev. E}, 80:056602, Nov 2009.

\bibitem{whitham2011linear}
Gerald~Beresford Whitham.
\newblock {\em Linear and nonlinear waves}.
\newblock John Wiley \& Sons, 2011.

\bibitem{Abeya_2023}
Asela Abeya, Gino Biondini, and Mark~A Hoefer.
\newblock Whitham modulation theory for the defocusing nonlinear schrödinger equation in two and three spatial dimensions.
\newblock {\em Journal of Physics A: Mathematical and Theoretical}, 56(2):025701, feb 2023.

\bibitem{https://doi.org/10.1111/sapm.12651}
Gino Biondini and Alexander Chernyavsky.
\newblock Whitham modulation theory for the zakharov–kuznetsov equation and stability analysis of its periodic traveling wave solutions.
\newblock {\em Studies in Applied Mathematics}, 152(2):596--617, 2024.

\bibitem{Ablowitz_2018}
Mark~J Ablowitz, Gino Biondini, and Igor Rumanov.
\newblock Whitham modulation theory for (2 + 1)-dimensional equations of kadomtsev–petviashvili type.
\newblock {\em Journal of Physics A: Mathematical and Theoretical}, 51(21):215501, apr 2018.

\bibitem{PhysRevE.96.032225}
Mark Ablowitz, Gino Biondini, and Qiao Wang.
\newblock Whitham modulation theory for the two-dimensional benjamin-ono equation.
\newblock {\em Phys. Rev. E}, 96:032225, Sep 2017.

\bibitem{EL201611}
G.A. El and M.A. Hoefer.
\newblock Dispersive shock waves and modulation theory.
\newblock {\em Physica D: Nonlinear Phenomena}, 333:11--65, 2016.
\newblock Dispersive Hydrodynamics.

\bibitem{carretero2024nonlinear}
R.~Carretero-Gonz{\'a}lez, D.J. Frantzeskakis, and P.G. Kevrekidis.
\newblock {\em Nonlinear Waves and Hamiltonian Systems: From One to Many Degrees of Freedom, from Discrete to Continuum}.
\newblock Oxford University Press, 2024.

\bibitem{evans1998partial}
L.C. Evans.
\newblock {\em Partial Differential Equations}.
\newblock Graduate studies in mathematics. American Mathematical Society, 1998.

\bibitem{https://doi.org/10.1111/sapm.12767}
Patrick Sprenger, Christopher Chong, Emmanuel Okyere, Michael Herrmann, P.~G. Kevrekidis, and Mark~A. Hoefer.
\newblock Hydrodynamics of a discrete conservation law.
\newblock {\em Studies in Applied Mathematics}, 153(4):e12767, 2024.

\bibitem{LAX1986250}
Peter~D. Lax.
\newblock On dispersive difference schemes.
\newblock {\em Physica D: Nonlinear Phenomena}, 18(1):250--254, 1986.

\bibitem{KAC1975160}
M~Kac and Pierre {van Moerbeke}.
\newblock On an explicitly soluble system of nonlinear differential equations related to certain toda lattices.
\newblock {\em Advances in Mathematics}, 16(2):160--169, 1975.

\bibitem{10.1063/1.1947120}
G.~A. El.
\newblock Resolution of a shock in hyperbolic systems modified by weak dispersion.
\newblock {\em Chaos: An Interdisciplinary Journal of Nonlinear Science}, 15(3):037103, 10 2005.

\end{thebibliography}

\bibliographystyle{unsrt}

\end{document}